\documentclass[apj]{emulateapj}

\def\ltsima{$\; \buildrel < \over \sim \;$}
\def\simlt{\lower.5ex\hbox{\ltsima}}
\def\gtsima{$\; \buildrel > \over \sim \;$}
\def\simgt{\lower.5ex\hbox{\gtsima}}
\def\kms{{\rm\,km\,s^{-1}}}
\def\kmsdeg{{\rm\,km\,s^{-1}\,deg^{-1}}}
\def\kmskpc{{\rm\,km\,s^{-1}\,kpc^{-1}}}

\def\kpc{{\rm\,kpc}}
\def\kpcdeg{{\rm\,kpc\,deg^{-1}}}

\def\msun{{\rm\,M_\odot}}
\def\lsun{{\rm\,L_\odot}}

\def\pc{{\rm\,pc}}

\def\deg{^\circ}

\def\s{\ifmmode \widetilde \else \~\fi}
\def\={\overline}

\def\spose#1{\hbox to 0pt{#1\hss}}

\def\lta{\mathrel{\spose{\lower 3pt\hbox{$\mathchar"218$}}
     \raise 2.0pt\hbox{$\mathchar"13C$}}}
\def\gta{\mathrel{\spose{\lower 3pt\hbox{$\mathchar"218$}}
     \raise 2.0pt\hbox{$\mathchar"13E$}}}
\def\Dt{\spose{\raise 1.5ex\hbox{\hskip3pt$\mathchar"201$}}}    
\def\dt{\spose{\raise 1.0ex\hbox{\hskip2pt$\mathchar"201$}}}    

\def\dotsfill{\leaders\hbox to 1em{\hss.\hss}\hfill}

\def\Gyr{{\rm\,Gyr}}
\def\FeH{{\rm[Fe/H]}}

\newcommand{\ud}{\mathrm{d}}

\shorttitle{}
\shortauthors{N. F. Martin \& S. Jin}

\begin{document}


\title{The Hercules satellite: a stellar stream in the Milky Way halo?}


\author{Nicolas F. Martin$^1$ \& Shoko Jin$^{2,\dagger}$}
\email{martin@mpia.de, shoko@ari.uni-heidelberg.de}

\altaffiltext{1}{Max-Planck-Institut f\"ur Astronomie, K\"onigstuhl 17, D-69117 Heidelberg, Germany}
\altaffiltext{2}{Astronomisches Rechen-Institut, Zentrum f\"ur Astronomie der Universit\"at Heidelberg, M\"onchhofstr. 12--14, D-69120, Heidelberg, Germany}
\altaffiltext{$\dagger$}{Alexander von Humboldt research fellow}

\begin{abstract}
We investigate the possibility that the recently discovered Hercules Milky Way satellite is in fact a stellar stream in formation, thereby explaining its very elongated shape with an axis ratio of 3 to 1. Under the assumption that Hercules is a stellar stream and that its stars are flowing along the orbit of its progenitor, we find an orbit that would have recently brought the system close enough to the Milky Way to induce its disruption and transformation from a bound dwarf galaxy into a stellar stream. The application of simple analytical techniques to the tentative radial velocity gradient observed in the satellite provides tight constraints on the tangential velocity of the system ($v_\mathrm{t} = -16^{+6}_{-22} \kms$ in the Galactic Standard of Rest). Combined with its large receding velocity, the determined tangential velocity yields an orbit with a small pericentric distance ($R_\mathrm{peri} = 6^{+9}_{-2} \kpc$). Tidal disruption is therefore a valid scenario for explaining the extreme shape of Hercules. The increase in the mean flattening of dwarf galaxies as one considers fainter systems could therefore be the impact of a few of these satellites not being bound stellar systems dominated by dark matter but, in fact, stellar streams in formation, shedding their stars in the Milky Way's stellar halo.
\end{abstract}


\section{Introduction}
The recent discoveries of numerous faint stellar systems around the Milky Way (MW; see e.g. \citealt{martin08b} and references therein) has thoroughly changed our view of its satellite system. The relatively high velocity dispersions measured from individual stars in these objects are usually seen as the sign that they are highly dark matter dominated \citep[e.g.][]{martin07a,simon07} and that they could play a significant role in explaining the apparent discrepancy between the number of dark matter subhalos seen in $\Lambda$CDM simulations compared to observed luminous dwarf galaxies in our surroundings \citep[e.g.][]{tollerud08,koposov09,maccio10}.

In spite of an ever-increasing knowledge of the properties of these newly discovered systems, their origin and nature still remain subject to interpretation. Are they `simply' faint equivalents of the previously known dwarf galaxies such as Draco, Sculptor or Fornax, leading to crucial constraints on galaxy formation and evolution? Are they disrupted/disrupting versions of formerly brighter progenitors that suffered a destructive fate from tidal interactions with the Milky Way? Or do they constitute a population of previously unknown systems, shaped by different formation and evolution mechanisms, that has remained hidden until the advent of systematic searches based on large surveys of the night sky?

The last scenario appears the least likely, given that the properties of the faint satellites are a continuous extension of our knowledge of Local Group dwarf galaxies: their sizes and central surface brightnesses are comparable, although they also extend to smaller/fainter scales \citep{martin08b}; they appear to globally follow the metallicity-luminosity relation followed by brighter galaxies \citep{kirby08}; they appear to inhabit similar dark matter halos \citep[e.g.,][]{strigari08,walker09}. Consequently, it seems more natural to envision them as extremely faint dwarf galaxies, as dwarf galaxy remnants, or as a combination of both. However, if one is to assume that all recently discovered systems are dwarf galaxies, \citet{martin08b} have shown that the faint MW satellite dwarf galaxies are significantly flatter than the brighter ones (observed mean ellipticity of $\langle\epsilon\rangle=0.47\pm0.03$ vs. $\langle\epsilon\rangle=0.32\pm0.02$).

In a thorough analysis of the observational consequences of the tidal interaction of a dwarf galaxy with the MW, \citet{munoz08} show that there is only a transient increase in the ellipticity of the satellite as it interacts with its host. They nevertheless show that, when the satellite is in the final throes of its destruction, it can show ellipticities as high as $\sim0.7$. In this case, the system has lost more than 90\% of its stars and is becoming unbound, thereby dissolving into a stellar stream. In this context, it is interesting to note that the large ellipticity measured for systems with $M_V\gta-7.5$ is mainly driven by three satellites (about a quarter of the sample) that are among the most flattened of all Galactic satellites: Ursa Major~I (UMaI; $\epsilon=0.80\pm0.04$), Hercules ($\epsilon=0.68^{+0.06}_{-0.08}$) and Ursa Major~II (UMaII; $\epsilon=0.63^{+0.03}_{-0.05}$). These values are reminiscent of those measured by \citet{munoz08} and suggest that these three systems could be transforming into stellar streams after their last, destructive, pericentric passage that brought them too close to the MW.

The morphology of these systems is also unlike what is seen for rounder dwarf galaxies: deep photometric follow-up observations confirm them all to be very elongated and with somewhat distorted morphologies \citep[][although see \citealt{martin08b} for the large impact of noise on their distorted shape]{coleman07,okamoto08,munoz10}. This is consistent with stellar systems becoming unbound, as studied by \citet{kroupa97}\footnote{See, for instance, the comparison of the map of Kroupa's simulation with that of Hercules obtained by \citet{coleman07}, as presented in \citet{kroupa10}.}. UMaII even shows a power-law radial density profile, which is typical of a disrupting system \citep[e.g.][]{johnston99,penarrubia09}. Regardless of their shape and structure, the possibility that these systems are transforming into streams is also tied to them having a pericentric distance that is small enough for them to be destroyed by the MW's tidal forces. If this could easily be the case for UMaII that is currently at a heliocentric distance of $\sim30\kpc$ \citep{zucker06b}, the viability of an orbit with a small pericenter needs to be investigated further for UMaI and Hercules, residing at distances of $97\pm4\kpc$ \citep{okamoto08} and $138\pm7\kpc$ (from an error-weighted averaging of the values from \citealt{aden09a} and \citealt{sand09}), respectively. UMaI unfortunately has a small radial velocity with respect to the Galactic Standard of Rest\footnote{Unless specified otherwise, we use Galactic Standard of Rest velocities throughout this paper; that is, velocities observed from the position of the Sun but corrected for the motion of the Sun around the MW \citep{dehnen98}.} ($v_\mathrm{r}  \simeq-10\kms$; \citealt{kleyna05,martin07a,simon07}), which places it close to its pericenter or apocenter and limits the constraints that one can place on its orbit. Hercules, on the other hand, has a large receding velocity ($v_\mathrm{r}=145\kms$; \citealt{simon07,aden09b}) that could be the consequence of a very radial orbit.

In this paper, we solve for the orbit of Hercules with the assumption that it is a stellar stream, in other words a disrupting dwarf galaxy that is no longer bound. Our goal is to verify that, under these conditions, there is indeed a viable orbit for the system; that is, an orbit that brings it close to the MW and can therefore explain its transformation into a stream. We use simple dynamical arguments based on the observation of a tentative radial velocity gradient in the system \citep{aden09b} to show that there is only a restricted range of tangential velocities --- and consequently orbits --- allowed for Hercules if it is a stellar stream. The pericenter of Hercules is very small, thereby confirming that this satellite could well be an unbound stellar system. The paper is organized as follows: in Section~2, we determine the properties of the orbit of Hercules under the hypothesis that it is a stream, Section~3 discusses our findings and investigates possible discrepancies between a Hercules stellar stream and current observations of the system, while Section~4 concludes this work.

\section{Determining the orbit of the Hercules stream}
\begin{figure}
\begin{center}
\includegraphics[width=\hsize]{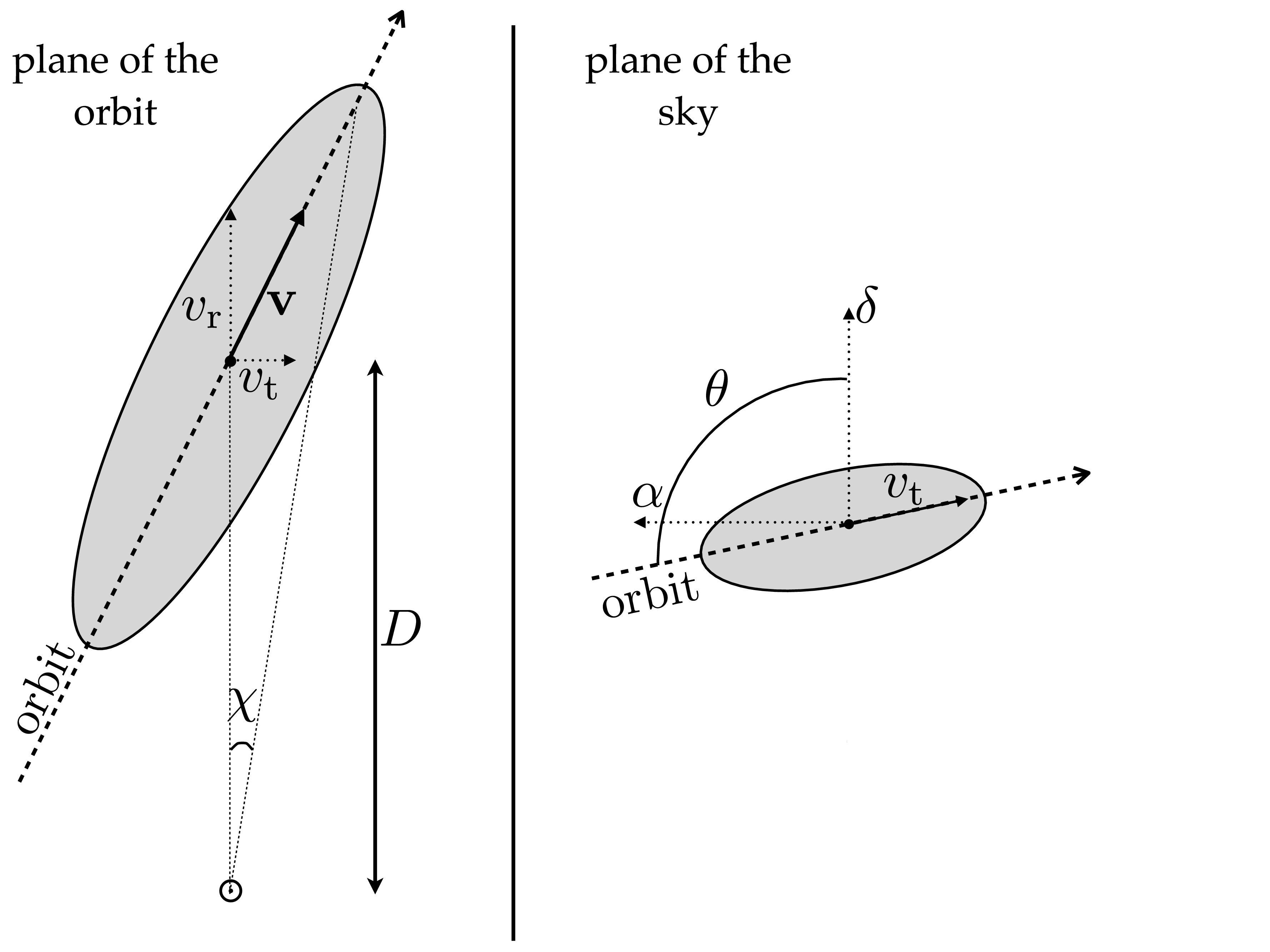}
\caption{\label{sketch}Sketch of the Hercules system in the plane of the orbit (left) and in the plane of the sky (right). The variables we use in this paper are indicated: the velocity vector ${\bf v}$, decomposed into the radial velocity, $v_\mathrm{r}$, and the tangential velocity of the system, $v_\mathrm{t}$; the distance to the center of Hercules, $D$; the angular distance along the orbit, measured from the center of Hercules, $\chi$; and its position angle, $\theta$, defined from equatorial North to East. In the plane of the sky, North is to the top and East to the left.}
\end{center}
\end{figure}

The basic assumption of this paper is that Hercules is no longer a dwarf galaxy, but instead a disrupted stellar system that is transforming into a stream. In this case, the system has clearly been stripped of all of its dark matter and its stars do not form a bound system anymore but are instead freely streaming along the orbit of its progenitor\footnote{Although the stars of a stripped system do not exactly follow the orbit of the progenitor, this simplification has little impact on our analysis.}. One thus expects a radial velocity gradient along the major axis of the system that \citet{jin07} and \citet{jin08th} have formalized in the case of generic orbits (see also \citealt{binney08b}). We refer the reader to these aforementioned sources for more detail but, in summary, if we are to follow the naming conventions shown in the sketch of Figure~\ref{sketch} and denote $\chi$ to be the angle along the orbit (measured from the center of Hercules and increasing towards decreasing right ascension), $D$ the heliocentric distance to Hercules, $v_\mathrm{r}$ and $v_\mathrm{t}$ the radial and tangential velocities of the orbit in the Galactic Standard of Rest, and $\nabla_\mathrm{r}\psi$ the gradient of the potential along the line of sight, then the radial velocity gradient along Hercules, $\ud v_\mathrm{r}/\ud\chi$, is a solution to the following quadratic equation:

\begin{equation}
\label{eqn:radvelrun}
\frac{\ud v_\mathrm{r}}{\ud\chi} = v_\mathrm{t} + \left(\nabla_\mathrm{r}\psi\right)\frac{D}{v_\mathrm{t}}.
\end{equation}

\noindent This results in the following two tangential velocity solutions of the orbit:

\begin{equation}
\label{eqn:vt}
v_\mathrm{t} = \frac{1}{2} \left( \frac{\ud v_\mathrm{r}}{\ud\chi} \pm \sqrt{\left(\frac{\ud v_\mathrm{r}}{\ud\chi}\right)^2 - 4\left(\nabla_\mathrm{r}\psi\right)D}\right).
\end{equation}

Since the distance $D$ to Hercules is known and the radial gradient of the potential $\nabla_\mathrm{r}\psi$ can be modeled, a direct measurement of a velocity gradient over the body of the satellite translates to only two possible orbits, corresponding to the positive and negative solutions of $v_\mathrm{t}$. 

\subsection{Modeling the velocity distribution of Hercules stars}
With this direct relationship between the radial velocity gradient and the orbit of Hercules in mind, we revisit the determination of a radial velocity gradient presented by \citet{aden09b}. From a sample of 18 carefully selected Hercules member stars, these authors show the tentative presence of a radial velocity gradient in a direction roughly consistent with the major axis of the system.

\subsubsection{The model}
We employ a maximum likelihood algorithm to fit a velocity gradient through the same data points they have used. Our model has four parameters: the radial velocity gradient, $\ud v_\mathrm{r}/\ud\chi$, the mean radial velocity at the center of Hercules $\overline{v_\mathrm{r}}$, the direction of the linear gradient on the sky, defined by its position angle from equatorial North to East, $\theta$, and the velocity dispersion, $s$, of member stars around the velocity gradient. The goal of the maximum likelihood technique is to find the set of these four parameters that maximizes the likelihood function 

\begin{equation}
\mathcal{L}\left(\frac{\ud v_\mathrm{r}}{\ud\chi},\overline{v_\mathrm{r}},\theta,s\right) = \prod_i \ell_i\left(\frac{\ud v_\mathrm{r}}{\ud\chi},\overline{v_\mathrm{r}},\theta,s\right),
\end{equation}

\noindent where $\ell_i\left(\ud v_\mathrm{r}/\ud\chi,\overline{v_\mathrm{r}},\theta,s\right)$ is the probability of finding the datum $i$ given the set of parameters. In the current problem, each star is defined by its radial velocity, $v_{\mathrm{r},i}$, and associated uncertainty, $v_{\mathrm{err},i}$, as well as its right ascension and declination $(\alpha_i,\delta_i)$ that we convert to its distance from the center of Hercules $(\alpha_0,\delta_0)$ such that $X_i = (\alpha_i-\alpha_0)\cos(\delta_0)$ and $Y_i = \delta_i-\delta_0$. The angular distance of star $i$ along an axis of position angle $\theta$ is then $y_i = X_i \sin(\theta) +Y_i\cos(\theta)$, yielding a difference between the modeled and measured radial velocities at this position of

\begin{equation}
\Delta v_{\mathrm{r},i} = v_{\mathrm{r},i} - \left(\frac{\ud v_\mathrm{r}}{\ud\chi}y_i + \overline{v_\mathrm{r}}\right).
\end{equation}

\noindent This velocity difference around the radial velocity gradient is finally modeled by a Gaussian, whose standard deviation is the velocity dispersion of the system, $s$, added in quadrature to the uncertainty in the velocity measurement of star $i$. This leads to the following expression for $\ell_i$:

\begin{equation}
\ell_i\left(\frac{\ud v_\mathrm{r}}{\ud\chi},\overline{v_\mathrm{r}},\theta,s\right) = \frac{1}{\sqrt{2\pi\left(s^2+v_{\mathrm{err},i}^2\right)}}\exp\left(-\frac{1}{2}\frac{\Delta v_{\mathrm{r},i}^2}{s^2+v_{\mathrm{err},i}^2}\right).
\end{equation}

Now that the likelihood function is entirely expressed as functions of the parameters and the data point properties, the model that maximizes it is determined by exploring a fine grid over the four dimensions of the parameter space\footnote{One could wonder whether a model with a radial velocity gradient that has one more parameters than a model with a fixed mean velocity and velocity dispersion is warranted when only 18 stars are available. A likelihood-ratio test in fact yields that the simpler model can be rejected at the 94\% confidence-level.}. In order to determine the uncertainties on the measurement of a given parameter $p_j$, we marginalize over the other three parameters and assume that the marginalized likelihood function $\mathcal{L'}$ is well-behaved near the best model (i.e, Gaussian-like or close to Gaussian). This allows us to use the property that $2\ln(\mathcal{L'})$ behaves as a $\chi^2$ distribution and determine the $k$-$\sigma$ confidence interval of parameter $p_j$ as being bound by the values of $p_j$ that correspond to $2\ln(\mathcal{L'})$ dropping by $k^2$. The uncertainties given below correspond to this definition of the 1-$\sigma$ confidence interval.

\subsubsection{Results}
We have checked that, if we are to force no velocity dispersion for the model ($s=0\kms$), the best model of \citet{aden09b} is recovered with a heliocentric $\overline{v_\mathrm{r}} = 45.1 \pm 0.4\kms$, and $\ud v_\mathrm{r}/\ud\chi = 15.4 \pm 2.8 \kmskpc$ for a position angle of $\theta=-35\deg$. If all the parameters are allowed to evolve freely, they converge on the following best values: heliocentric $\overline{v_\mathrm{r}} = 45.0 \pm 1.1\kms$, $s = 3.5^{+1.2}_{-0.9} \kms$ and $\ud v_\mathrm{r}/\ud\chi = 14.1^{+7.7}_{-7.3} \kmskpc$ (as part of the velocity gradient is accounted for by the assumed internal velocity dispersion), for a position angle of $\theta=-37\deg$.

\begin{figure}
\begin{center}
\includegraphics[width=0.87\hsize,angle=270]{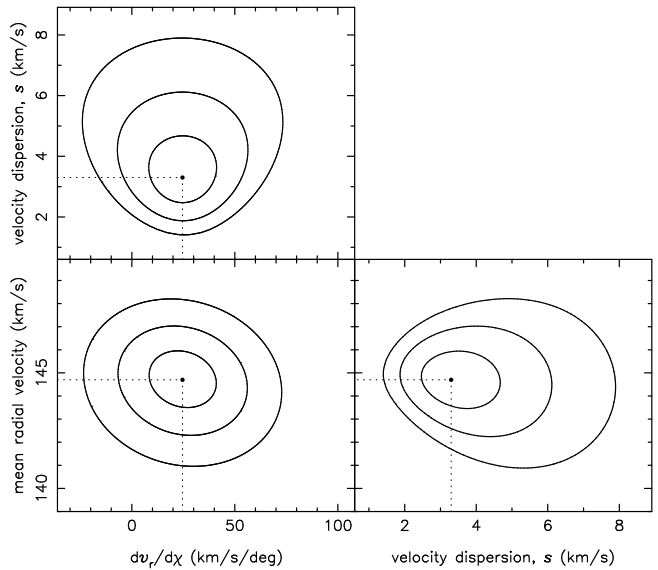}
\caption{\label{2dlike}Two-dimensional likelihood contours for parameters of our model. The filled circle represents the best model for each parameter set.  From this point outwards, the contours indicate drops in likelihood of 50\%, 90\% and 99\%. These are very regular, showing that the maxima of the likelihood distributions are well defined.}
\end{center}
\end{figure}

However, under our assumption that Hercules is a stellar stream, the axis of the radial velocity gradient must be the axis along which the system is elongated, that is, its major axis. Consequently, the position angle of the system, $\theta=-78\pm4\deg$ \citep{martin08b}, is used as a prior in order to find the best model, whose two-dimensional marginalized likelihood functions are shown in Figure~\ref{2dlike} for the three remaining parameters. From the one-dimensional marginalized likelihood functions, we determine the following values for the best model: $\overline{v_\mathrm{r}} = 144.7 \pm 1.2\kms$, $s = 3.5^{+1.1}_{-0.9} \kms$ and $\ud v_\mathrm{r}/\ud\chi = 10.2 \pm 6.0 \kmskpc = 24.6 \pm 14.5 \kmsdeg$.

\subsection{The distribution of likely Hercules orbits}
\label{orbit}
\begin{figure*}
\begin{center}
\includegraphics[width=0.32\hsize,angle=270]{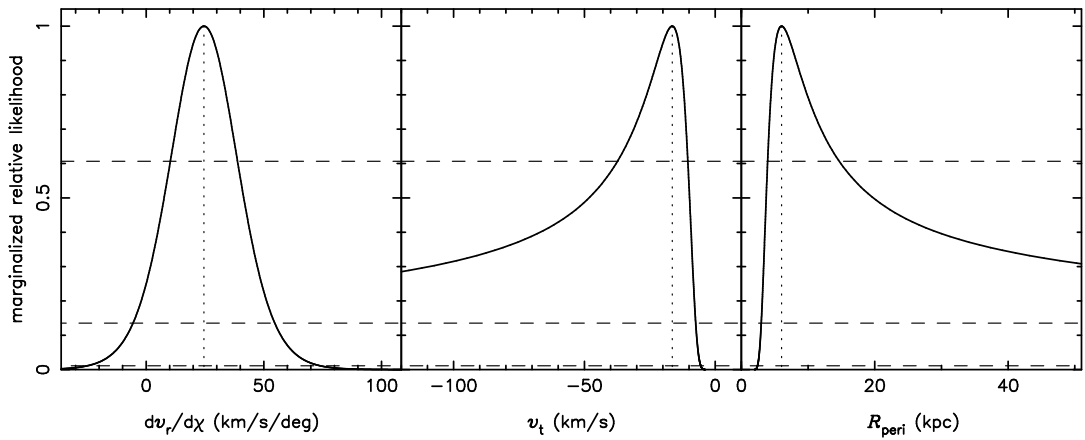}
\caption{\label{1dlike}Marginalized relative likelihood distributions for the radial velocity gradient, tangential velocity and perigalacticon, whose best values are $\mathrm{d}v_\mathrm{r}/\mathrm{d}\chi = 25\pm15\kmsdeg$, $v_\mathrm{t} = -16^{+6}_{-22}\kms$ and $R_\mathrm{peri} = 6^{+9}_{-2}\kpc$, respectively.  From top to bottom, the dashed lines intersect the likelihood functions at the boundaries of the 1-, 2- and 3-$\sigma$ confidence intervals.}
\end{center}
\end{figure*}

Having obtained the likelihood distribution of the velocity gradient, shown in its marginalized form in the left panel of Figure~\ref{1dlike}, it is now possible to use equation~(\ref{eqn:vt}) to determine the likelihood distribution of the tangential velocity of the Hercules orbit. The distance is assumed to be $D=138\pm7\kpc$, obtained by averaging the distance measurements of \citet{aden09a} and \citet{sand09}. The Milky Way potential we place ourselves in is a mixture model which combines the Miyamoto-Nagai disk and bulge defined by \citet{paczynski90} with the adiabatically contracted NFW halo constrained by \citet{xue08}, leading to $\nabla_\mathrm{r}\psi=-170(\kms)^2\kpc^{-1}$ at the location of Hercules. Choosing the positive solution for $v_\mathrm{t}$ yields very large velocities that produce unbound orbits for Hercules with a very large pericenter. This is clearly not a viable solution for a system disrupted by a recent passage close to the Milky Way center. The negative solution for $v_\mathrm{t}$, on the other hand, yields much more interesting results. The marginalized likelihood distribution for this case is shown in the middle panel of Figure~\ref{1dlike} and reveals a preferred velocity of $v_\mathrm{t} = -16^{+6}_{-22} \kms$, consistent with a very radial orbit given the mean velocity of Hercules ($\overline{v_\mathrm{r}} = 144.7 \pm 1.2\kms$) at its center.

\begin{figure}
\begin{center}
\includegraphics[width=0.87\hsize]{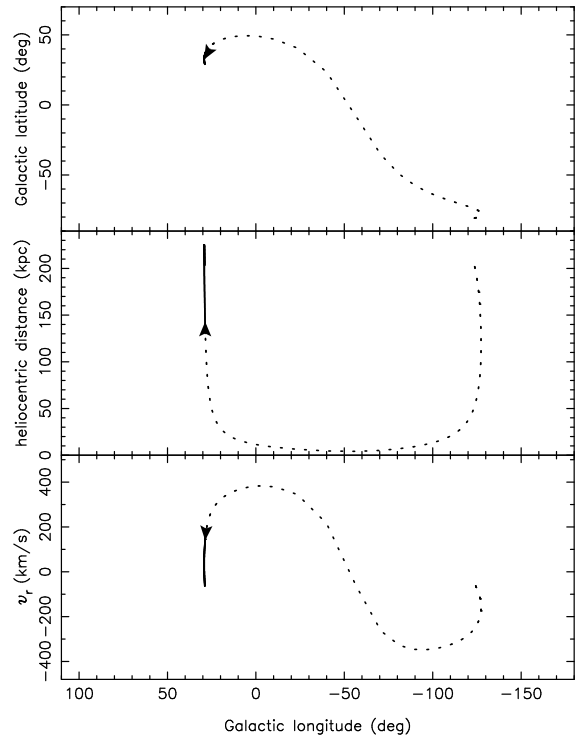}
\caption{\label{lbdv}Orbit of the Hercules stellar stream using our best model as given in the text, integrated forward and backward from the present location for $2\Gyr$ in each direction.  From top to bottom, the panels show the location of the orbit on the sky in Galactic coordinates, as well as its heliocentric distance and radial velocity relative to the Galactic Standard of Rest.  The solid and dotted lines respectively denote the forward and backward-integrated orbit from the present location of Hercules, which is indicated by an arrowhead in the direction of the orbit.}
\end{center}
\end{figure}

Given that the sky position, radial velocity, distance to and direction of motion of the assumed Hercules stream are known, determining the tangential velocity likelihood distribution of the system is equivalent to knowing its orbit likelihood distribution. In particular, there is a direct relationship between $\mathcal{L'}(v_\mathrm{t})$ and the likelihood distribution of the stream's Galactocentric distance at the last pericenter, $\mathcal{L'}(R_\mathrm{peri})$. By integrating the orbits for 2\,Gyr backward and forward in time in the potential described above, we obtain the distribution of $\mathcal{L'}(R_\mathrm{peri})$, also shown in Figure~\ref{1dlike}, from which we determine tight constraints on the small pericenter of the orbit: $R_\mathrm{peri}=6^{+9}_{-2}\kpc$. This pericenter was reached $\sim0.6\Gyr$ ago. The properties of the corresponding preferred orbit are shown in Figure~\ref{lbdv}.

\section{Discussion}
Under the assumption that the radial velocity gradient observed along the body of Hercules is due to the system being a stellar stream, we are able to find an orbit that has brought it to within $6^{+9}_{-2}\kpc$ of the Galactic center. But is this close enough to induce the destruction of a dwarf galaxy into a stellar stream during the last $0.6\Gyr$ and does the very radial orbit, with a perigalacticon to apogalacticon ratio of 6:219 (i.e. an eccentricity $e=0.95$), contradict the current knowledge of Hercules? We now discuss these two points and the various consequences of the orbit we have derived above on the observational properties of the stellar system.

\subsection{A small enough pericenter for tidal disruption?}
Whether or not a pericenter of $6^{+9}_{-2}\kpc$ is small enough to induce the tidal destruction of Hercules is a function of the properties of the stellar system at that time of its journey in the Milky Way potential and, as such, out of reach to the current observer. We can however use observed properties of Hercules to conservatively estimate the minimum pericentric radius it should have reached to undergo a strong tidal interaction with the Milky Way.

By definition \citep{king62}, at its pericenter $R_\mathrm{peri}$, a satellite is pruned to its tidal radius, $r_t$, such that

\begin{equation}
\label{rt}
r_t\simeq R_\mathrm{peri} \left(\frac{m}{M(R_\mathrm{peri})(3+e)}\right)^{1/3},
\end{equation}

\noindent where $e$ is the eccentricity of the satellite's orbit around the Galaxy, $m$ is the total mass of the satellite and $M(R_\mathrm{peri})$ is the mass of the Milky Way within the pericentric radius. Assuming a constant circular velocity $v_c$ (expressed in $\kms$) for the plausible range of Hercules pericentric distances, this mass can be expressed (with $R_\mathrm{peri}$ expressed in \kpc) as

\begin{equation}
\label{mass}
M(R_\mathrm{peri})\simeq\frac{R_\mathrm{peri} v_c^2}{4}\times10^6 \msun.
\end{equation}

\noindent Finally, if we assume that $v_c\simeq200\kms$ \citep{xue08}, inserting equation~(\ref{mass}) into equation~(\ref{rt}) and solve for $R_\mathrm{peri}$, we obtain

\begin{equation}
\label{peri}
R_\mathrm{peri} \simeq r_t^{3/2} \left( \frac{(3+e)}{\,m}\times10^{10}\right)^{1/2},
\end{equation}

\noindent with $m$ expressed in solar units.

We are left with a function of the properties of Hercules at pericenter and we know that, for the favored orbit determined above, $e=0.95$. The two remaining unknowns need to be assumed. In the case of the tidal radius at pericenter, advocating for tidal disruption requires it to be much smaller than the current extent of the system that has an on-sky King limiting radius of $1.4\kpc$ \citep{sand09}. We assume $r_t\simeq0.3\kpc$, a value that is close to the current half-light radius of Hercules \citep{martin08b,sand09}, as a necessary condition for the system to be stripped of a large fraction of its stars and transformed into stellar stream. Regarding the mass, we use the value determined by \citet{aden09b} within 300\pc: $m\simeq2\times10^6\msun$. This leads to $R_\mathrm{peri}\simeq23\kpc$ being required for a strong interaction to take place between Hercules and the Milky Way. This value is of course very uncertain as it is strongly dependent on our choice of $r_t$ and $m$ but it confirms that, with $R_\mathrm{peri}=6^{+9}_{-2}\kpc$, the preferred orbit determined in Section~\ref{orbit} appears to have a small enough pericenter to induce the transformation of Hercules into a stellar stream.

\subsection{The shape of Hercules}
With the distance to Hercules being much larger than the Sun-to-Galactic-center distance, the system's orbit being very radial means that its physical, three-dimensional size is much larger than its projected size as measured on the plane of sky. Transforming the measured half-light radius and ellipticity measured on the sky, $r_h$ and $\epsilon=1-b/a$, into their de-projected equivalents, $r_{h\mathrm{, deproj}}$ and $\epsilon_\mathrm{deproj}$, can be done easily from simple geometrical considerations obtained directly from Figure~\ref{sketch}:

\begin{eqnarray}
r_{h\mathrm{, deproj}} & = & r_h (\cos\chi)^{-1} =  r_h \left(1+\frac{v_\mathrm{r}^2}{v_\mathrm{t}^2}\right)^{1/2}\\
\epsilon_\mathrm{deproj} & = & 1-\frac{b}{a}\left(1+\frac{v_\mathrm{r}^2}{v_\mathrm{t}^2}\right)^{-1/2}.
\end{eqnarray}

From on-sky measurements ($r_h = 230\pm30\pc$ and $\epsilon=0.67\pm0.03$; \citealt{sand09}), these relations yield a de-projected half-light radius and ellipticity of $r_{h\mathrm{, deproj}}= 1.5\kpc$ and $\epsilon_\mathrm{deproj}=0.95$. One should note, however, that in the case of a stellar stream, the half-light radius is not a very meaningful quantity as the stream's surface brightness profile is not expected to follow the typical exponential, Plummer or King profiles that have so far been used to derive the properties of this satellite. Disrupting systems are expected to follow power-law density profiles in their outskirts (e.g. \citealt{johnston99,penarrubia09}), but current data have not yet been used to quantify the validity of such a density model for Hercules.

\subsection{The issue of the distance gradient}
If Hercules is a stellar stream as opposed to a dwarf galaxy, its stars are, as mentioned before, freely streaming along its orbit. This has the consequence of naturally producing the radial velocity gradient that was used in Section~2 to constrain the orbit, but it also produces a distance gradient along the body of the system that is all the more important as the orbit is very radial. In \citet{jin09}, we have studied this distance gradient in the generic case and shown that, following the notations used here, it can be expressed as:

\begin{equation}
\label{eqn:vs_dist}
\frac{\ud D}{\ud\chi} = D \frac{v_\mathrm{r}}{v_\mathrm{t}}.
\end{equation}

\noindent With the properties of the Hercules orbit determined above, it follows that the system should exhibit a distance gradient of $\ud D/\ud \chi = -22^{+8}_{-30}\kpcdeg$. 

The current best estimate of a distance gradient, based on color-magnitude diagram (CMD) fitting of small regions on either side of Hercules' center, has been determined by \citet{sand09}. They find that the distance gradient over the system is at most of $\sim6\kpc$ over $\sim6$\,arcmin, or $\ud D/\ud\chi \simeq 60\kpcdeg$. This gradient goes in the opposite direction to the one we determine from our orbit, but \citet{sand09} show that their detection of a distance gradient is very tentative. In fact, from their Figures~13 and~14, one can see that it is just as likely for an inverse gradient to be present in their data. Thus, there is no outstanding discrepancy between the current observations of the system and the distance gradient that should be observed if Hercules follows the orbit we propose in this paper. However, given the low density of Hercules' stars, it will be difficult to observe the predicted distance gradient. Contrary to, for instance, Draco (e.g. \citealt{klessen03}), the system hosts but a few horizontal branch stars from which an accurate distance gradient measurement could be derived and the alternative of using deep data reaching the main sequence turn-off is plagued by contamination at the faint end of the CMD (cf. the \citealt{sand09} analysis). Even though the orbit of the Hercules stream is very radial, confirming the presence of its large distance gradient will be a trying task. A search for the few RR Lyrae stars that should inhabit the system might be the best alley of investigation to ever constrain the presence of a distance gradient.

\subsection{A tidally disrupted Hercules?}
If there is no evidence that Hercules cannot be a stream in formation, there are in fact some signs that it could not be a pristine, bound stellar system. The deepest available data of Hercules \citep{coleman07,sand09} show the presence of some material along the major axis of the system, extending away from its main body. \citet{sand09} highlight that this extension has a CMD that is tentatively similar to that of Hercules and also coincides with a small overdensity of likely Hercules blue horizontal-branch stars. In addition, further along the major axis (or likely orbit of the stream), they investigate the nature of three small clumps of stars that appear in their smoothed maps within $\sim0.5\deg$ of Hercules' centroid, and again find tentative evidence of a connection to the stellar system. 

\citet{aden09b} also find some asymmetry in the distribution of their carefully selected Hercules member red giant branch stars that they ascribe to the possible effect of tides. Although this extension of three stars is not aligned with the major axis and therefore seems discrepant with our scenario of Hercules being a stellar stream, one has to note that their photometric analysis only covers a limited field of view that does not favor the detection of stars extending along the major axis of the system. Thus, it is not inconsistent with the scenario we propose.

In addition, and as mentioned in the introduction, Hercules has a striking resemblance with the result of the \citet{kroupa97} simulations which emphasize that, even though they are not systems in equilibrium at the center of massive dark matter halos, disrupting stellar systems can have properties consistent with those observed for dwarf galaxies. A recent, more detailed re-simulation of this initial work by \citet{metz07} further shows that such disrupting systems can share the observed properties of the population of recently discovered, faint, MW dwarf galaxies. There are therefore ways to form a disrupting stellar system that shares the observed properties of Hercules.

This interpretation seems to contradict that of \citet{penarrubia08b}, who find that the recently discovered satellites, including Hercules, are unlikely to be the remnants of brighter MW dwarf galaxies that once shared the current properties of Draco, Fornax, or Sagittarius. Their conclusion mainly stems from the relatively large velocity dispersions of the newly discovered stellar systems. \citet{aden09a} have, however, shown that Hercules in fact has a low velocity dispersion that we find to be even smaller when one properly accounts for the observed velocity gradient in the system. Following backwards the `tidal evolutionary tracks' measured by \citet{penarrubia08b} to track the evolution of the properties of observed dwarf galaxies embedded in $\Lambda$CDM dark matter halos as they undergo tidal disruption (their Figure~10), the current velocity dispersion and luminosity  of Hercules $(s,L)=(3.5\kms,3.6\times10^4\lsun)$ correspond to the properties expected for an initial satellite with $(s,L)=(8.7\kms,3.6\times10^5\lsun)$ that has been stripped of 90\% of its stellar mass\footnote{This $\sim90\%$ loss in stellar mass is a necessary condition for the dwarf galaxies simulated by \citet{munoz08} to become unbound and show large ellipticities.}. These properties are in fact very similar to the current velocity dispersion and luminosity of the Draco or Ursa Minor dwarf galaxies, making it possible that Hercules is actually the remnant of such a galaxy.

However, if this is truly what is happening to Hercules, the low metallicity measured for its stars (mean $\FeH$ between $-2.6$ and $-2.0$, depending on the study; \citealt{simon07,kirby08,koch08b,aden09a}) and its broad agreement with the luminosity/metallicity relation followed by dwarf galaxies \citep{kirby08} would seem to imply that the progenitor of the observed system could not have been much brighter than the current luminosity of Hercules ($M_V\simeq-6.5$; \citealt{martin08b,sand09}). In fact, the large spread in the relation would nevertheless make it possible for the observed stellar system to be a disrupted version of Draco or Ursa Minor ($\FeH\simeq-2.0$; \citealt{winnick03}). A violent tidal interaction with the Milky Way, which is consistent with the very eccentric orbit we have determined ($e=0.95$), would also allow for a system which quickly goes from bound to disrupted, without the need for a long, gradual peeling off of its stars that would allow the progenitor to stay bound for a longer period of time.

There remains the issue of timing and how likely it is to be observing a system that is exactly in its disruption phase but that is still concentrated enough for it to be observed as an overdensity of stars. This could make for an uncomfortable coincidence but, as has been previously mentioned by \citet{aden09b}, \citet{fellhauer07} have shown in simulations of the UMaII stellar system that this process requires $\sim1$\,Gyr. With the last, destructive pericenter of the orbit we have determined happening only $\sim0.6\Gyr$ ago, it would therefore not be surprising that we could still be observing the core of a Hercules system currently transforming into a stream. The UMaII simulation with its different orbit is of course not directly comparable to the Hercules case, and detailed simulations of the evolution of a disrupting Hercules following the orbit we have determined are warranted, but we currently find no outstanding issues with the scenario taken as the assumption of this paper.

\subsection{Hercules mass estimates}
By removing foreground Milky Way stars that happen to share velocities similar to that of member stars, the velocity dispersion of Hercules and hence its mass have been severely corrected downwards to yield a dynamical mass within $300\pc$ of only $M_{300}=1.9^{+1.1}_{-0.8}\times10^6\msun$ \citep{aden09b}. This value is low enough that it falls off the \citet{strigari08} `common mass-scale' of dwarf galaxies that all share  $M_{300}\sim10^7\msun$. This scale can easily be explained by dwarf galaxy formation models (e.g. \citealt{maccio08,li09,okamoto09}) and the Hercules dwarf galaxy being a significant outlier could be worrying.

One must nevertheless remember that there are numerous assumptions that enter the determination of such dynamical masses, first and foremost that the stellar system is in virial equilibrium. This is obviously not a valid assumption if the system is currently disrupting and this would lead to spurious mass estimates. Taken in conjunction with the tentative observation of a velocity gradient and the presence of some stars possibly streaming out of the system, the abnormally low velocity dispersion of Hercules could be yet another sign that it is in fact a stream in formation, as opposed to a peculiar bound dwarf galaxy.

\section{Conclusion}

We have shown that, under the assumption that the tentative velocity gradient observed in Hercules is a sign that it is no longer a bound dwarf galaxy but instead a stellar stream in formation, there is a viable orbit that can explain the properties of the system. We constrain the tangential velocity of Hercules to be only $v_\mathrm{t}=-16^{+6}_{-22}\kms$, which makes for a very eccentric orbit ($e=0.95$) whose pericentric distance is only $R_\mathrm{peri}=6^{+9}_{-2} \kpc$. This value is small enough that it could explain the tidal disruption of a satellite that we currently see as a stellar stream in formation after its last destructive pericentric passage close to the MW center.

We do not claim that there can be no other scenario that could explain the observed properties of the system and the hypothesis of a stream in formation that we have presented in this paper may not necessarily be the plight of Hercules. Our intention is to provide avenues of investigation in order to understand the peculiar properties of some of the recently discovered faint dwarf galaxies. We show that, in the case of Hercules, tidal destruction seems to be a perfectly viable option to explain its surprisingly large ellipticity and low velocity dispersion. It is worth emphasizing that we are \emph{not} advocating that \emph{all} faint dwarf galaxies are stellar streams but simply that \emph{some} systems, namely those with very large ellipticities (e.g. Hercules, UMaI, UMaII), could well be shaped by such a process. This hypothesis can, and should, be tested by deeper and wider photometric follow-up that can reveal these systems to be much larger than originally measured (e.g. \citealt{munoz10} for UMaII) and/or, in the case of Hercules, by more extended spectroscopic coverage that would definitely confirm (or rule out) the presence of the radial velocity gradient that is our core assumption.

If the systems with the largest measured ellipticities are indeed stellar streams in formation, that would alleviate the necessity of having to find a galaxy formation mechanism that produces more flattened systems at the faint end of the luminosity function. This simple possibility that Hercules could be an unbound system should also make one wary of treating the population of newly discovered stellar systems as a single population of objects.

\acknowledgments
It is a pleasure to thank Jelte de Jong, Jorge Pe\~narrubia and, in particular, Hans-Walter Rix for stimulating discussions. NFM would also like to thank the organizers and visitors of the Kavli Institute for Theoretical Physics program `Back to the Galaxy II', during which part of this work was performed.


\begin{thebibliography}{39}
\expandafter\ifx\csname natexlab\endcsname\relax\def\natexlab#1{#1}\fi

\bibitem[{{Ad{\'e}n} {et~al.}(2009{\natexlab{a}}){Ad{\'e}n}, {Feltzing},
  {Koch}, {Wilkinson}, {Grebel}, {Lundstr{\"o}m}, {Gilmore}, {Zucker},
  {Belokurov}, {Evans}, \& {Faria}}]{aden09a}
{Ad{\'e}n}, D., {Feltzing}, S., {Koch}, A., {Wilkinson}, M.~I., {Grebel},
  E.~K., {Lundstr{\"o}m}, I., {Gilmore}, G.~F., {Zucker}, D.~B., {Belokurov},
  V., {Evans}, N.~W., \& {Faria}, D. 2009{\natexlab{a}}, \aap, 506, 1147

\bibitem[{{Ad{\'e}n} {et~al.}(2009{\natexlab{b}}){Ad{\'e}n}, {Wilkinson},
  {Read}, {Feltzing}, {Koch}, {Gilmore}, {Grebel}, \&
  {Lundstr{\"o}m}}]{aden09b}
{Ad{\'e}n}, D., {Wilkinson}, M.~I., {Read}, J.~I., {Feltzing}, S., {Koch}, A.,
  {Gilmore}, G.~F., {Grebel}, E.~K., \& {Lundstr{\"o}m}, I. 2009{\natexlab{b}},
  \apjl, 706, L150

\bibitem[{{Binney}(2008)}]{binney08b}
{Binney}, J. 2008, \mnras, 386, L47

\bibitem[{{Coleman} {et~al.}(2007){Coleman}, {de Jong}, {Martin}, {Rix},
  {Sand}, {Bell}, {Pogge}, {Thompson}, {Hippelein}, {Giallongo}, {Ragazzoni},
  {DiPaola}, {Farinato}, {Smareglia}, {Testa}, {Bechtold}, {Hill}, {Garnavich},
  \& {Green}}]{coleman07}
{Coleman}, M.~G., {de Jong}, J.~T.~A., {Martin}, N.~F., {Rix}, H.-W., {Sand},
  D.~J., {Bell}, E.~F., {Pogge}, R.~W., {Thompson}, D.~J., {Hippelein}, H.,
  {Giallongo}, E., {Ragazzoni}, R., {DiPaola}, A., {Farinato}, J., {Smareglia},
  R., {Testa}, V., {Bechtold}, J., {Hill}, J.~M., {Garnavich}, P.~M., \&
  {Green}, R.~F. 2007, \apjl, 668, L43

\bibitem[{{Dehnen} \& {Binney}(1998)}]{dehnen98}
{Dehnen}, W., \& {Binney}, J. 1998, \mnras, 294, 429

\bibitem[{{Fellhauer} {et~al.}(2007){Fellhauer}, {Evans}, {Belokurov},
  {Zucker}, {Yanny}, {Wilkinson}, {Gilmore}, {Irwin}, {Bramich}, {Vidrih},
  {Hewett}, \& {Beers}}]{fellhauer07}
{Fellhauer}, M., {Evans}, N.~W., {Belokurov}, V., {Zucker}, D.~B., {Yanny}, B.,
  {Wilkinson}, M.~I., {Gilmore}, G., {Irwin}, M.~J., {Bramich}, D.~M.,
  {Vidrih}, S., {Hewett}, P., \& {Beers}, T. 2007, \mnras, 375, 1171

\bibitem[{{Jin}(2008)}]{jin08th}
{Jin}, S. 2008, PhD thesis, Cambridge University

\bibitem[{{Jin} \& {Lynden-Bell}(2007)}]{jin07}
{Jin}, S., \& {Lynden-Bell}, D. 2007, \mnras, 378, L64

\bibitem[{{Jin} \& {Martin}(2009)}]{jin09}
{Jin}, S., \& {Martin}, N.~F. 2009, \mnras, 400, L43

\bibitem[{{Johnston} {et~al.}(1999){Johnston}, {Sigurdsson}, \&
  {Hernquist}}]{johnston99}
{Johnston}, K.~V., {Sigurdsson}, S., \& {Hernquist}, L. 1999, \mnras, 302, 771

\bibitem[{{King}(1962)}]{king62}
{King}, I. 1962, \aj, 67, 471

\bibitem[{{Kirby} {et~al.}(2008){Kirby}, {Simon}, {Geha}, {Guhathakurta}, \&
  {Frebel}}]{kirby08}
{Kirby}, E.~N., {Simon}, J.~D., {Geha}, M., {Guhathakurta}, P., \& {Frebel}, A.
  2008, \apjl, 685, L43

\bibitem[{{Klessen} {et~al.}(2003){Klessen}, {Grebel}, \&
  {Harbeck}}]{klessen03}
{Klessen}, R.~S., {Grebel}, E.~K., \& {Harbeck}, D. 2003, \apj, 589, 798

\bibitem[{{Kleyna} {et~al.}(2005){Kleyna}, {Wilkinson}, {Evans}, \&
  {Gilmore}}]{kleyna05}
{Kleyna}, J.~T., {Wilkinson}, M.~I., {Evans}, N.~W., \& {Gilmore}, G. 2005,
  \apjl, 630, L141

\bibitem[{{Koch} {et~al.}(2008){Koch}, {McWilliam}, {Grebel}, {Zucker}, \&
  {Belokurov}}]{koch08b}
{Koch}, A., {McWilliam}, A., {Grebel}, E.~K., {Zucker}, D.~B., \& {Belokurov},
  V. 2008, \apjl, 688, L13

\bibitem[{{Koposov} {et~al.}(2009){Koposov}, {Yoo}, {Rix}, {Weinberg},
  {Macci{\`o}}, \& {Escud{\'e}}}]{koposov09}
{Koposov}, S.~E., {Yoo}, J., {Rix}, H.-W., {Weinberg}, D.~H., {Macci{\`o}},
  A.~V., \& {Escud{\'e}}, J.~M. 2009, \apj, 696, 2179

\bibitem[{{Kroupa}(1997)}]{kroupa97}
{Kroupa}, P. 1997, New Astronomy, 2, 139

\bibitem[{{Kroupa} {et~al.}(2010){Kroupa}, {Famaey}, {de Boer},
  {Dabringhausen}, {Pawlowski}, {Boily}, {Jerjen}, {Forbes}, {Hensler}, {Del
  Popolo}, \& {Metz}}]{kroupa10}
{Kroupa}, P., {Famaey}, B., {de Boer}, K.~S., {Dabringhausen}, J., {Pawlowski},
  M.~S., {Boily}, C.~M., {Jerjen}, H., {Forbes}, D., {Hensler}, G., {Del
  Popolo}, A., \& {Metz}, M. 2010, ArXiv:1006.1647K

\bibitem[{{Li} {et~al.}(2009){Li}, {Helmi}, {De Lucia}, \& {Stoehr}}]{li09}
{Li}, Y., {Helmi}, A., {De Lucia}, G., \& {Stoehr}, F. 2009, \mnras, 397, L87

\bibitem[{{Macci{\`o}} {et~al.}(2008){Macci{\`o}}, {Dutton}, \& {van den
  Bosch}}]{maccio08}
{Macci{\`o}}, A.~V., {Dutton}, A.~A., \& {van den Bosch}, F.~C. 2008, \mnras,
  391, 1940

\bibitem[{{Macci{\`o}} {et~al.}(2010){Macci{\`o}}, {Kang}, {Fontanot},
  {Somerville}, {Koposov}, \& {Monaco}}]{maccio10}
{Macci{\`o}}, A.~V., {Kang}, X., {Fontanot}, F., {Somerville}, R.~S.,
  {Koposov}, S., \& {Monaco}, P. 2010, \mnras, 402, 1995

\bibitem[{{Martin} {et~al.}(2008){Martin}, {de Jong}, \& {Rix}}]{martin08b}
{Martin}, N.~F., {de Jong}, J.~T.~A., \& {Rix}, H.-W. 2008, \apj, 684, 1075

\bibitem[{{Martin} {et~al.}(2007){Martin}, {Ibata}, {Chapman}, {Irwin}, \&
  {Lewis}}]{martin07a}
{Martin}, N.~F., {Ibata}, R.~A., {Chapman}, S.~C., {Irwin}, M., \& {Lewis},
  G.~F. 2007, \mnras, 380, 281

\bibitem[{{Metz} \& {Kroupa}(2007)}]{metz07}
{Metz}, M., \& {Kroupa}, P. 2007, \mnras, 376, 387

\bibitem[{{Mu{\~n}oz} {et~al.}(2010){Mu{\~n}oz}, {Geha}, \&
  {Willman}}]{munoz10}
{Mu{\~n}oz}, R.~R., {Geha}, M., \& {Willman}, B. 2010, \aj, 140, 138

\bibitem[{{Mu{\~n}oz} {et~al.}(2008){Mu{\~n}oz}, {Majewski}, \&
  {Johnston}}]{munoz08}
{Mu{\~n}oz}, R.~R., {Majewski}, S.~R., \& {Johnston}, K.~V. 2008, \apj, 679,
  346

\bibitem[{{Okamoto} {et~al.}(2008){Okamoto}, {Arimoto}, {Yamada}, \&
  {Onodera}}]{okamoto08}
{Okamoto}, S., {Arimoto}, N., {Yamada}, Y., \& {Onodera}, M. 2008, \aap, 487,
  103

\bibitem[{{Okamoto} \& {Frenk}(2009)}]{okamoto09}
{Okamoto}, T., \& {Frenk}, C.~S. 2009, \mnras, 399, L174

\bibitem[{{Paczy\'nski}(1990)}]{paczynski90}
{Paczy\'nski}, B. 1990, \apj, 348, 485

\bibitem[{{Pe{\~n}arrubia} {et~al.}(2008){Pe{\~n}arrubia}, {Navarro}, \&
  {McConnachie}}]{penarrubia08b}
{Pe{\~n}arrubia}, J., {Navarro}, J.~F., \& {McConnachie}, A.~W. 2008, \apj,
  673, 226

\bibitem[{{Pe{\~n}arrubia} {et~al.}(2009){Pe{\~n}arrubia}, {Navarro},
  {McConnachie}, \& {Martin}}]{penarrubia09}
{Pe{\~n}arrubia}, J., {Navarro}, J.~F., {McConnachie}, A.~W., \& {Martin},
  N.~F. 2009, \apj, 698, 222

\bibitem[{{Sand} {et~al.}(2009){Sand}, {Olszewski}, {Willman}, {Zaritsky},
  {Seth}, {Harris}, {Piatek}, \& {Saha}}]{sand09}
{Sand}, D.~J., {Olszewski}, E.~W., {Willman}, B., {Zaritsky}, D., {Seth}, A.,
  {Harris}, J., {Piatek}, S., \& {Saha}, A. 2009, \apj, 704, 898

\bibitem[{{Simon} \& {Geha}(2007)}]{simon07}
{Simon}, J.~D., \& {Geha}, M. 2007, \apj, 670, 313

\bibitem[{{Strigari} {et~al.}(2008){Strigari}, {Bullock}, {Kaplinghat},
  {Simon}, {Geha}, {Willman}, \& {Walker}}]{strigari08}
{Strigari}, L.~E., {Bullock}, J.~S., {Kaplinghat}, M., {Simon}, J.~D., {Geha},
  M., {Willman}, B., \& {Walker}, M.~G. 2008, \nat, 454, 1096

\bibitem[{{Tollerud} {et~al.}(2008){Tollerud}, {Bullock}, {Strigari}, \&
  {Willman}}]{tollerud08}
{Tollerud}, E.~J., {Bullock}, J.~S., {Strigari}, L.~E., \& {Willman}, B. 2008,
  \apj, 688, 277

\bibitem[{{Walker} {et~al.}(2009){Walker}, {Mateo}, {Olszewski},
  {Pe{\~n}arrubia}, {Wyn Evans}, \& {Gilmore}}]{walker09}
{Walker}, M.~G., {Mateo}, M., {Olszewski}, E.~W., {Pe{\~n}arrubia}, J., {Wyn
  Evans}, N., \& {Gilmore}, G. 2009, \apj, 704, 1274

\bibitem[{{Winnick}(2003)}]{winnick03}
{Winnick}, R.~A. 2003, PhD thesis, Yale University

\bibitem[{{Xue} {et~al.}(2008){Xue}, {Rix}, {Zhao}, {Re Fiorentin}, {Naab},
  {Steinmetz}, {van den Bosch}, {Beers}, {Lee}, {Bell}, {Rockosi}, {Yanny},
  {Newberg}, {Wilhelm}, {Kang}, {Smith}, \& {Schneider}}]{xue08}
{Xue}, X.~X., {Rix}, H.~W., {Zhao}, G., {Re Fiorentin}, P., {Naab}, T.,
  {Steinmetz}, M., {van den Bosch}, F.~C., {Beers}, T.~C., {Lee}, Y.~S.,
  {Bell}, E.~F., {Rockosi}, C., {Yanny}, B., {Newberg}, H., {Wilhelm}, R.,
  {Kang}, X., {Smith}, M.~C., \& {Schneider}, D.~P. 2008, \apj, 684, 1143

\bibitem[{{Zucker} {et~al.}(2006){Zucker}, {Belokurov}, {Evans}, {Kleyna},
  {Irwin}, {Wilkinson}, {Fellhauer}, {Bramich}, {Gilmore}, {Newberg}, {Yanny},
  {Smith}, {Hewett}, {Bell}, {Rix}, {Gnedin}, {Vidrih}, {Wyse}, {Willman},
  {Grebel}, {Schneider}, {Beers}, {Kniazev}, {Barentine}, {Brewington},
  {Brinkmann}, {Harvanek}, {Kleinman}, {Krzesinski}, {Long}, {Nitta}, \&
  {Snedden}}]{zucker06b}
{Zucker}, D.~B., {Belokurov}, V., {Evans}, N.~W., {Kleyna}, J.~T., {Irwin},
  M.~J., {Wilkinson}, M.~I., {Fellhauer}, M., {Bramich}, D.~M., {Gilmore}, G.,
  {Newberg}, H.~J., {Yanny}, B., {Smith}, J.~A., {Hewett}, P.~C., {Bell},
  E.~F., {Rix}, H.-W., {Gnedin}, O.~Y., {Vidrih}, S., {Wyse}, R.~F.~G.,
  {Willman}, B., {Grebel}, E.~K., {Schneider}, D.~P., {Beers}, T.~C.,
  {Kniazev}, A.~Y., {Barentine}, J.~C., {Brewington}, H., {Brinkmann}, J.,
  {Harvanek}, M., {Kleinman}, S.~J., {Krzesinski}, J., {Long}, D., {Nitta}, A.,
  \& {Snedden}, S.~A. 2006, \apjl, 650, L41

\end{thebibliography}

\clearpage

\end{document}